\documentclass[conference]{IEEEtran}

\usepackage{amsmath,amssymb}
\usepackage{booktabs}

\usepackage{cite}

\usepackage[xindy,toc]{glossaries}
\usepackage{graphicx} % Required for inserting images

\usepackage{multirow}

\usepackage[dvipsnames]{xcolor}

\newacronym{CPU}{CPU}{Central Processing Unit}

\newacronym{ILP}{ILP}{Integer Linear Programming}
\newacronym{MILP}{MILP}{Mixed-Integer Linear Programming}
\newacronym{RL}{RL}{Reinforcement Learning}
\newacronym{MARL}{MARL}{Multi Agent Reinforcement Learning}
\newacronym{AI}{AI}{Artificial Intelligence}
\newacronym{ML}{ML}{Machine Learning}
\newacronym{DRL}{DRL}{Deep Reinforcement Learning}
\newacronym{MDP}{MDP}{Markov Decision Process}
\newacronym{PPO}{PPO}{Proximal Policy Optimization}
\newacronym{LLM}{LLM}{Large Language Model}
\newacronym{MEC}{MEC}{Multi-access Edge Computing}
\newacronym{NFV}{NFV}{Network Function Virtualization}
\newacronym{VNF}{VNF}{Virtualized Network Function}
\newacronym{CNF}{CNF}{Cloud-Native Network Function}
\newacronym{5G}{5G}{5th Generation}
\newacronym{6G}{6G}{6th Generation}
\newacronym{SDN}{SDN}{Software-Defined Networking}
\newacronym{O-RAN}{O-RAN}{Open Radio Access Network}
\newacronym{RIC}{RIC}{RAN Intelligent Controller}
\newacronym{RT}{RT}{Real Time}
\newacronym{SMO}{SMO}{Service Management and Orchestration}
\newacronym{GoS}{GoS}{Grade of Service}

\newacronym{O-RU}{O-RU}{Open Radio Units}
\newacronym{O-DU}{O-DU}{Open Distributed Units}
\newacronym{O-CU}{O-CU}{Open Central Units}
\newacronym{UPF}{UPF}{User Plane Function}
\newacronym{QoS}{QoS}{Quality of Service}
\newacronym{UE}{UE}{User Equipment}
\newacronym{TN}{TN}{Transport Network}
\newacronym{CN}{CN}{Core Network}

\newacronym{CAPEX}{CAPEX}{Capital Expenditure}
\newacronym{SP}{SP}{Shortest-Path}
\newacronym{MinLat}{MinLat}{Minimum Latency}
\newacronym{MinEnergy}{MinEnergy}{Minimum Energy}

\newacronym{LT}{LT}{Linear Thompson}
\newacronym{BTE}{BTE}{Bootstrap Tree Ensemble}
\title{Objective-Aware Policy Orchestration for O-RAN Service Provisioning}

\author{\IEEEauthorblockN{Sebastian Racedo and Brigitte Jaumard}
\IEEEauthorblockA{\textit{Computer Science and Software Eng.} \\
\textit{Concordia University}\\
Montreal (Qc) Canada \\
brigitte.jaumard@concordia.ca}
\and
\IEEEauthorblockN{Meysam Masoudi}
\IEEEauthorblockA{\textit{KTH Royal Institute of Technology} \\
Stockholm, Sweden}
\and
\IEEEauthorblockN{Oscar Delgado}
\IEEEauthorblockA{\textit{Computer Science and Software Eng.} \\
\textit{Concordia University} \\
Montreal (Qc) Canada}
}

\begin{document}

\maketitle

\begin{abstract}
\acrshort{O-RAN} enables the deployment of heterogeneous control and provisioning algorithms through open interfaces and programmable intelligent controllers. 
However, most existing intelligent \acrshort{O-RAN} solutions rely on a single optimization strategy designed for a fixed operational objective, limiting their ability to adapt when network conditions or operator priorities change.
This paper proposes an objective-aware policy orchestration framework that selects the most appropriate provisioning policy for each control window based on the current network context and operator objective. 

The framework orchestrates a portfolio of heterogeneous provisioning experts using contextual-bandit learning. 
Two policy selectors, Linear Thompson Sampling and a Bootstrap Tree ensemble, 
are evaluated against static provisioning policies and a deterministic rule-based orchestrator. 
Experimental results show that learning-based policy orchestration can effectively adapt policy selection to varying operating conditions, improving overall objective-aware performance while preserving the modularity of existing provisioning algorithms.
\end{abstract}


% ----------------------------------------
% ----------------------------------------
% ----------------------------------------

\section{Introduction}
\label{sec:introduction}

The disaggregation and programmability of \gls{O-RAN} enable adaptive control through near-real-time applications for traffic steering, resource allocation, admission control, and service assurance, and non-real-time applications for policy management, model training, analytics, and longer-term optimization \cite{Polese_2023}. Operators can deploy heuristics,
optimization models, and learned policies, but must still determine which algorithm suits the current traffic, objectives, and resources.

Most intelligent \gls{O-RAN} studies focus on resource
orchestration, directly allocating spectrum, compute, and transport resources \cite{Staffolani_2024,Qiao_2025,Lotfi_2025}. Policies designed for a particular objective may not remain best as conditions change: latency-oriented experts may protect delay-sensitive services, energy-oriented experts may suit stable periods, and learned policies
may balance interacting constraints. Changing an \gls{RL} policy's objective usually requires reward changes and retraining, limiting operational flexibility.

This work studies a different decision layer: policy orchestration. Rather than directly selecting a route, a virtualized function location, or a resource allocation for every request, the proposed orchestrator selects one approved provisioning policy map for the next control window. The lower-level provisioning expert remains responsible for processing individual service-chain requests and enforcing bandwidth, latency, compute, and topology constraints. The orchestration layer therefore operates at a slower timescale and can be conceptually placed in the Non-\gls{RT} \gls{RIC} or the \gls{SMO} framework. The selected map is frozen during the target window, preventing per-request policy switching and separating high-level orchestration from low-level execution.

The orchestration decision depends on both the operator objective and the network context. The context includes the performance of the completed observation window, available compute and transport resources, congestion indicators, failure information, a target-window traffic forecast, and operational information such as the currently active policy. The operator objective indicates whether the next decision should prioritize \gls{GoS}, energy consumption, latency, congestion mitigation, or a balanced combination. Consequently, the same physical network state may lead to a different preferred expert when the operator objective changes.

The high-level action is selected using contextual bandits. This formulation is appropriate because the current problem is a finite, one-window decision: observe the context, choose a policy map, execute it for one window, and observe an objective-aware reward. Contextual bandits are more data-efficient than a high-level deep reinforcement-learning controller and do not require learning a multi-window transition model \cite{Monea_2024} . We consider two complementary learners. \gls{LT} Sampling provides regularized action-specific reward models and explicit uncertainty estimates, while a \gls{BTE} captures nonlinear interactions between objectives, traffic, failures, and resource conditions.

The contributions of this work are summarized as follows:
\begin{itemize}
    \item An objective-aware policy orchestration framework for \gls{O-RAN} service provisioning that dynamically selects among existing provisioning policies according to operator objectives, network conditions, and traffic context, enabling flexible adaptation without modifying the underlying experts.
    \item A comprehensive evaluation of the proposed orchestration framework under multiple operator objectives (e.g., \gls{GoS}, energy efficiency, and balanced operation), demonstrating that objective-aware policy selection provides a more adaptable alternative to static single-policy deployment.
\end{itemize}

The remainder of this paper is organized as follows. 
Section \ref{sec:related_work} briefly reviews previous work on orchestration in \gls{O-RAN} and 5G. 
We formally define the problem statement in Section \ref{sec:problem_statement}. 
Methodology is presented in Section \ref{sec:methodology}.
Numerical results are presented in Section \ref{sec:Results}.
Section \ref{sec:Conclusion} concludes the paper and discusses future work.

\section{Literature Review}
\label{sec:related_work}
\subsection{O-RAN Management and Resource Orchestration}
The \gls{O-RAN} architecture separates control functions according to their timescale and scope. The Near-\gls{RT} \gls{RIC} hosts xApps for control loops typically operating between tens of milliseconds and one second, whereas the Non-\gls{RT} \gls{RIC} and \gls{SMO} support longer-term analytics, policy management, model training, service orchestration, and lifecycle management \cite{Polese_2023}. \gls{AI}-native management frameworks extend this architecture by treating \gls{AI} models, data pipelines, and intelligent control functions as first-class network components \cite{Habibi_2024,Li_2025}.

A substantial portion of the literature applies \gls{AI} directly to resource-allocation problems. PRORL, for example, uses \gls{DRL} to anticipate demand and proactively allocate \gls{O-RAN} infrastructure resources \cite{Staffolani_2024}. Hierarchical learning has also been considered for network-slicing resource allocation, reflecting the natural separation between higher-level orchestration and lower-level control in \gls{O-RAN} \cite{Qiao_2025}. Related work employs \gls{DRL} for radio resource management, scheduling, admission control, and slice configuration. In these approaches, the learned agent is itself the resource-allocation policy.

Although such methods demonstrate the ability of \gls{RL} to model nonlinear resource interactions, they are commonly designed around a fixed reward or fixed multi-objective trade-off. Consequently, modifying the operator priority may require retraining or redesigning the low-level policy. In contrast, the problem studied here assumes that several provisioning experts already exist and asks which expert should be activated under the current objective and operating context.

\subsection{xApp, rApp, and Policy Orchestration}
A related research direction considers the management of intelligent applications themselves. OREO studies the deployment and sharing of xApps while accounting for service quality and resource consumption \cite{Puligheddu_2025}. Other frameworks investigate xApp and rApp conflicts, policy coordination, and application lifecycle management \cite{Brach_2025}. These works establish that operating several intelligent applications requires a decision layer above the applications themselves.

Hierarchical and intent-driven approaches are particularly related to the present work because a higher-level controller may select or coordinate lower-level xApps according to operator goals \cite{Mallu_2023, Habib_2023_Intent_driven}. However, existing solutions commonly coordinate applications belonging to the same algorithmic family or generate control configurations directly. Less attention has been given to selecting among heterogeneous provisioning experts—for example, deterministic shortest-path, latency-oriented, energy-oriented, and \gls{RL} policies—using a common objective-aware evaluation model.

The distinction between resource orchestration and policy orchestration is therefore central. Resource orchestration determines quantities such as bandwidth, compute, or power. Policy orchestration determines which control logic should make those lower-level decisions. The latter is the focus of this paper.

\subsection{LLMs and Agentic AI for Telecom Networks}
Recent work investigates \gls{LLM}s and agentic systems for intent interpretation, policy generation, knowledge retrieval, and autonomous network management. Domain-specific foundation models such as ORANSight explore natural-language interaction with \gls{O-RAN} knowledge and specifications \cite{Gajjar_2025}. \gls{LLM}-augmented \gls{RL} approaches use language models to provide contextual guidance or improve learning for network slicing \cite{Lotfi_2025}. AgentRAN proposes an agentic architecture in which distributed agents interpret operator intents and coordinate network control functions \cite{Elkael_2026}. Multi-agentic frameworks have also been proposed for conflict-aware rApp policy orchestration \cite{Li_2026_Multi_Agentic}.

These developments motivate the long-term integration of intent translation, planning, tool use, validation, execution, and feedback. Nevertheless, much of the emerging agentic literature remains conceptual, prepublication, or limited to controlled prototypes. In addition, an unconstrained \gls{LLM} should not directly modify a network configuration because generated actions may be infeasible or unsafe. A practical architecture should separate language-based intent interpretation from numerical policy evaluation, constraint verification, and controlled execution.

The current work focuses on the learning-based decision component that such an agentic architecture would require. Operator objectives are supplied in a structured form rather than generated by an \gls{LLM}, and the available actions are restricted to a validated policy catalog. This isolates the policy-selection problem and allows it to be evaluated independently of natural-language interpretation.

Existing studies address low-level resource allocation, application deployment, intent translation, and AI-native network management. Limited empirical work examines an intermediate, window-based orchestration layer that observes an operator objective together with network KPIs, forecasts, failures, and resource conditions; selects among heterogeneous deterministic and learned provisioning experts;  evaluates the decision with a common objective-aware reward; and learns the context in which each expert is preferable without retraining the experts themselves. This paper addresses that gap through an objective-conditioned contextual-bandit formulation.

\section{Problem Statement}
\label{sec:problem_statement}

Let the physical infrastructure be represented by a directed graph $\mathcal{G}=(\mathcal{V},\mathcal{E})$,
%\begin{equation}
%    \mathcal{G}=(\mathcal{V},\mathcal{E}),
%\end{equation}
where $\mathcal{V}$ contains radio, transport, compute, and core-network nodes, and $\mathcal{E}$ contains the transport links connecting them. Each link $e\in\mathcal{E}$ has bandwidth capacity $B_e$ and propagation delay $d_e$. Compute-capable nodes have finite CPU, memory, and storage capacities. The initial dimensioning of the network and the traffic used can be found in \cite{Racedo_CCECE_2026}.

Time is divided into control windows indexed by $t\in\mathcal{T}$. During window $t$, the set of service-chain requests is denoted by $\mathcal{Q}_t$. Each request $q\in\mathcal{Q}_t$ is characterized by a service class $s_q\in\mathcal{S}$, a bandwidth requirement $b_q$, a maximum admissible latency $\ell_q^{\max}$, and an ordered \gls{O-RAN} chain and core-network functions. Once a provisioning policy is activated, it is responsible for embedding/routing these requests subject to topology, bandwidth, latency, and compute-resource constraints.

Let $\Pi=\left\{\pi_{\mathrm{SP}}, \pi_{\mathrm{Lat}}, \pi_{\mathrm{Eng}}, \pi_{\mathrm{PPO}} \right\}$
%\begin{equation}
%    \Pi=
%    \left\{
%        \pi_{\mathrm{SP}},
%        \pi_{\mathrm{Lat}},
%        \pi_{\mathrm{Eng}},
%        \pi_{\mathrm{PPO}}
%    \right\}
%\end{equation}
denote the available provisioning experts, corresponding respectively to shortest-path, minimum-latency, minimum-energy, and learned PPO-based (see \cite{Racedo_CNSM_2025} for more details on the selected PPO agent) provisioning. The power model used to calculate the energy was taken from \cite{Racedo_FGCS_2026}. 

A policy map $a$ assigns one provisioning expert to each service class: $a:\mathcal{S}\rightarrow\Pi.$
%\begin{equation}
%    a:\mathcal{S}\rightarrow\Pi.
%\end{equation}
In the general formulation, different service classes may be assigned different experts. In the current implementation, however, the action catalog is restricted to four global policy maps: $\mathcal{A}= \left\{a_{\mathrm{SP}}, a_{\mathrm{Lat}}, a_{\mathrm{Eng}}, a_{\mathrm{PPO}}\right\},$
%\begin{equation}
%    \mathcal{A}=
%    \left\{
%        a_{\mathrm{SP}},
%        a_{\mathrm{Lat}},
%        a_{\mathrm{Eng}},
%        a_{\mathrm{PPO}}
%    \right\},
%\end{equation}
where each map assigns the same expert to every service class. This restriction isolates the high-level policy-selection problem from the more complex problem of generating service-specific policy maps. A common expert doesn't imply identical allocations: requests retain their service class characteristics. Consistent with NG-RAN's distinction between differentiated handling and resource isolation \cite{3GPP_TS38300_R17}, our model preserves service-specific constraints without establishing per-slice isolation or service-level guarantees. Concurrent service-specific expert selection remains future work.

Before target window $t$, the orchestrator observes an operational context $\mathbf{x}_t$ containing the active operator objective and the network information available before the decision. It then selects one eligible policy map $a_t\in\mathcal{A}_t\subseteq\mathcal{A},$
%\begin{equation}
%    a_t\in\mathcal{A}_t\subseteq\mathcal{A},
%\end{equation}
which remains fixed throughout the target window. 

Executing policy map $a$ under context $\mathbf{x}_t$ produces a network outcome $\mathbf{z}_t(a)$, whose quality is quantified through the objective-aware score $R_t(a|\mathbf{x}_t)$
%\begin{equation}
%    R_t(a|\mathbf{x}_t).
%\end{equation}
which reflects the resulting service performance, energy consumption, latency, congestion, constraint compliance, and operational costs according to the active operator objective. The orchestration problem is therefore $a_t^{*} = \arg\max_{a\in\mathcal{A}_t} \mathbb{E} \left[ R_t(a\mid\mathbf{x}_t)\right]$.
%\begin{equation}
%    a_t^{*}
%    =
%    \arg\max_{a\in\mathcal{A}_t}
%    \mathbb{E}
%    \left[
%        R_t(a\mid\mathbf{x}_t)
%    \right].
%    \label{eq:orchestration_problem}
%\end{equation}

The objective is not to optimize individual routing or resource-allocation decisions directly. Instead, the orchestrator determines which provisioning expert should govern the next control window based on the current network context and operator objective. The selected expert is then responsible for all request-level provisioning decisions during that window.

\section{Methodology}
\label{sec:methodology}
This section presents the proposed objective-aware orchestration framework. First, the overall orchestration workflow is introduced. The provisioning expert portfolio and operator objectives are then described, followed by the construction of the decision context and the objective-aware evaluation model. Finally, the contextual-bandit selectors are presented.

%%%%%%%%%%%%%%%%%%%%%%%%%%%%%%%%%%%%%%%%%%%%%%%%%%%%%%%%%%%%%%%%%%%%%%%%%%%%%%
\subsection{Overall Orchestration Framework}
\label{subsec:framework}
\begin{figure}
    \centering
    \includegraphics[width=0.7\linewidth]{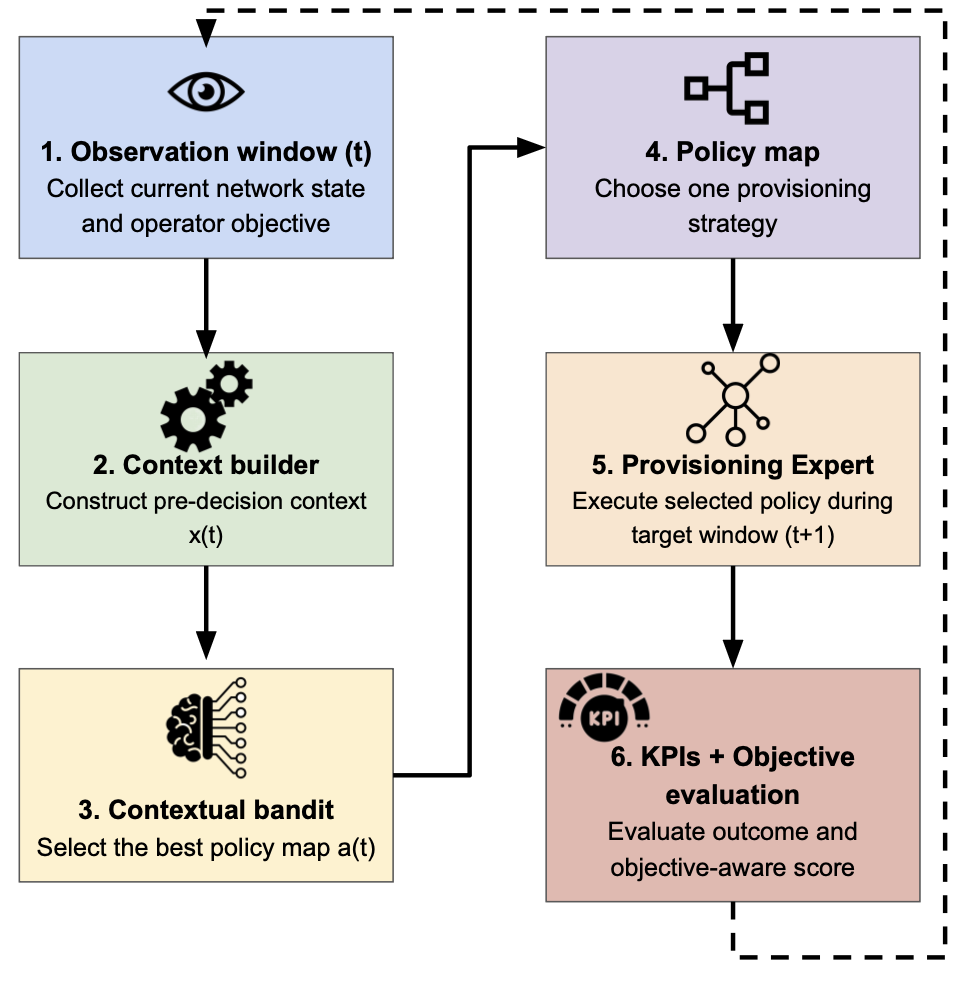}
    \caption{Overall Orchestration Framework}
    \label{fig:framework}
\end{figure}

Figure \ref{fig:framework} illustrates the proposed orchestration architecture.
Unlike conventional provisioning algorithms that directly determine routes or resource allocations for individual requests, the proposed framework operates at the policy level. At the beginning of every control window, the orchestrator observes the current network conditions together with the operator objective and selects one provisioning policy map from a predefined portfolio. The selected policy remains active during the entire target window, while all request-level provisioning decisions are delegated to the chosen expert. After the target window finishes, the resulting network performance is evaluated according to the active operator objective. Outcomes are recorded as feedback; the reported evaluation uses canonical one-step contexts rather than a selected-action closed loop.

The proposed architecture separates the orchestration process into four logical stages: \textit{i)} observation of the current network state, \textit{ii)} selection of the provisioning policy, \textit{iii)} execution of the selected provisioning expert, \textit{iv)} and objective-aware evaluation of the resulting network performance.

%\begin{enumerate}
%    \item observation of the current network state;
%    \item selection of the provisioning policy;
%    \item execution of the selected provisioning expert;
%    \item objective-aware evaluation of the resulting network performance.
%\end{enumerate}

%%%%%%%%%%%%%%%%%%%%%%%%%%%%%%%%%%%%%%%%%%%%%%%%%%%%%%%%%%%%%%%%%%%%%%%%%%%%%%
\subsection{Provisioning Expert Portfolio}
\label{subsec:experts}

The orchestrator selects one policy map from the expert portfolio introduced in Section~\ref{sec:problem_statement}. The portfolio contains four heterogeneous provisioning strategies: \gls{SP}, \gls{MinLat}, \gls{MinEnergy}, and PPO-based learned provisioning.
%\begin{itemize}
%    \item \gls{SP},
%    \item Minimum Latency (MinLat),
%    \item Minimum Energy (MinEnergy),
%    \item PPO-based learned provisioning.
%\end{itemize}
Each expert implements a complete provisioning policy capable of embedding and routing all service-chain requests during one control window. The orchestrator does not modify these experts or their internal optimization procedures. Instead, it learns under which operating conditions each expert is expected to provide the highest performance score. This separation allows new provisioning experts to be incorporated into the portfolio without modifying the orchestration framework.

%%%%%%%%%%%%%%%%%%%%%%%%%%%%%%%%%%%%%%%%%%%%%%%%%%%%%%%%%%%%%%%%%%%%%%%%%%%%%%
\subsection{Operator Objectives}
\label{subsec:objectives}

The proposed framework is objective-aware, meaning that the preferred provisioning policy depends on the operational objective specified by the network operator. Each objective consists of three conceptual components: a primary optimization objective, optional operational constraints, and secondary performance preferences.
%\begin{itemize}
%    \item a primary optimization objective;
%    \item optional operational constraints;
%    \item secondary performance preferences.
%\end{itemize}
The current implementation considers five representative operational modes: \textit{i)} \gls{GoS}, \textit{ii)} energy, \textit{iii)} latency, \textit{iv)} congestion, and \textit{v)} balanced. Consequently, identical network conditions may produce different policy selections depending on the active objective.

%%%%%%%%%%%%%%%%%%%%%%%%%%%%%%%%%%%%%%%%%%%%%%%%%%%%%%%%%%%%%%%%%%%%%%%%%%%%%%
\subsection{Pre-Decision Context Construction}
\label{subsec:context}
Before each target control window, the orchestrator constructs a context vector 
$\mathbf{x}_t = 
\left[ 
    \mathbf{x}^{\text{obj}}, \mathbf{x}^{\textsc{kpi}}, 
    \mathbf{x}^{\text{res}}, \mathbf{x}^{\text{forecast}}, \mathbf{x}^{\text{op}} 
\right],$
%\begin{equation}
%\mathbf{x}_t=
%\left[
%\mathbf{x}^{obj},
%\mathbf{x}^{kpi},
%\mathbf{x}^{res},
%\mathbf{x}^{forecast},
%\mathbf{x}^{op}
%\right],
%\end{equation}
using only information available prior to the decision. The context contains five groups of information: operator objective, completed-window KPIs, network and compute-resource utilization, traffic forecast for the next control window, and operational information such as the currently active policy map.
%\begin{itemize}
%    \item operator objective;
%    \item completed-window KPIs;
%    \item network and compute-resource utilization;
%    \item traffic forecast for the next control window;
%    \item operational information such as the currently active policy map.
%\end{itemize}
No information generated during the target window is included in the decision context, thereby preventing target leakage during training and evaluation.

%%%%%%%%%%%%%%%%%%%%%%%%%%%%%%%%%%%%%%%%%%%%%%%%%%%%%%%%%%%%%%%%%%%%%%%%%%%%%%
\subsection{Objective-Aware Outcome Evaluation}
\label{subsec:reward}
Let $G,E,L,u,c,f$ denote window GoS, energy (Wh),
average latency (ms), maximum link utilization, maximum CPU
utilization, and request failure rate, respectively.
Using $[x]_0^1=\min(1,\max(0,x))$, define
$v_G=[G]_0^1$, $v_E=[1-E/E_{\mathrm{ref}}]_0^1$,
$v_L=[1-L/L_{\mathrm{ref}}]_0^1$, $v_u=[1-u]_0^1$,
$v_c=[1-c]_0^1$, and $v_f=[1-f]_0^1$,
with fixed positive normalization bounds
$E_{\mathrm{ref}}$ (Wh) and $L_{\mathrm{ref}}$ (ms).
The primary score $S_{\mathrm{prim}}$ equals
$v_G$, $v_E$, $v_L$, or $v_u$ for GoS, energy, latency,
or congestion objectives, respectively; balanced operation
uses $\mathbf{w}_{\mathrm{B}}^\top\mathbf{v}$, where
$\mathbf{v}=(v_G,v_E,v_L,v_u,v_c,v_f)$.
Without explicit secondary priorities,
$U_{\mathrm{sec}}=\mathbf{w}_o^\top\mathbf{v}$,
where $\mathbf{w}_o$ contains objective-dependent,
nonnegative weights summing to one over these six KPIs.
Non-balanced modes exclude the primary KPI from the
secondary score; balanced mode includes all six.
Compliance requires a completed window, execution status
\texttt{success} or \texttt{degraded}, and satisfaction of
all configured hard KPI constraints; missing mandatory
metrics count as violations.
For compliant outcomes,
$R_t=[0.75S_{\mathrm{prim}}+0.25U_{\mathrm{sec}}
-C_{\mathrm{sw}}-C_{\mathrm{rec}}]_0^1$,
where $C_{\mathrm{sw}}=0.03\delta_t$,
$C_{\mathrm{rec}}=0.02\delta_t$, and
$\delta_t\in\{0,1\}$ indicates whether activation changes the current policy map. Otherwise, $R_t=-1+0.9f_{\mathrm{cons}}$, where
$f_{\mathrm{cons}}$ is the number of satisfied hard checks
divided by the total number evaluated.
Thus, compliant and noncompliant outcomes occupy
$[0,1]$ and $[-1,-0.1]$, respectively, preventing secondary
performance gains from outweighing realized constraint
violations.

%%%%%%%%%%%%%%%%%%%%%%%%%%%%%%%%%%%%%%%%%%%%%%%%%%%%%%%%%%%%%%%%%%%%%%%%%%%%%%
\subsection{Data Generation}
\label{subsec:dataset}
For every decision context, all candidate policy maps are independently executed from identical initial network conditions. Each execution produces one complete provisioning outcome while preserving the same observation context. This procedure generates a dataset of tuples $(\mathbf{x}_t,a,\mathbf{z}_t,R_t),$
%\begin{equation}
%(\mathbf{x}_t,a,\mathbf{z}_t,R_t),
%\end{equation}
where $\mathbf{x}_t$ is the pre-decision context, $a$ is the evaluated policy map, $\mathbf{z}_t$ denotes the resulting network outcome, and $R_t$ is the corresponding objective-aware score.
%\begin{itemize}
%    \item $\mathbf{x}_t$ is the pre-decision context,
%    \item $a$ is the evaluated policy map,
%    \item $\mathbf{z}_t$ denotes the resulting network outcome,
%    \item $R_t$ is the corresponding objective-aware score.
%\end{itemize}
Since every policy is evaluated from the same initial state, differences in performance can be directly attributed to the selected provisioning strategy rather than differences in network evolution.

%%%%%%%%%%%%%%%%%%%%%%%%%%%%%%%%%%%%%%%%%%%%%%%%%%%%%%%%%%%%%%%%%%%%%%%%%%%%%%
\subsection{Contextual Policy Selection}
\label{subsec:learners}
The policy-selection problem is formulated as a contextual-bandit task. Given the current context $\mathbf{x}_t$, the selector estimates the expected score of every eligible policy map $\hat{R}(a|\mathbf{x}_t),$
%\begin{equation}
%\hat{R}(a|\mathbf{x}_t),
%\end{equation}
and selects $a_t = \arg\max\limits_{a\in\mathcal{A}_t} \hat{R}(a|\mathbf{x}_t).$
%\begin{equation}
%a_t=
%\arg\max_{a\in\mathcal{A}_t}
%\hat{R}(a|\mathbf{x}_t).
%\end{equation}
Two complementary contextual-bandit models are investigated.

%%%%%%%%%%%%%%%%%%%%%%%%%%%%%%%%%%
\subsubsection{Linear Thompson (LT) Sampling}
The first learner models the expected score of every policy map using Bayesian linear regression. Thompson Sampling balances exploration and exploitation by sampling model parameters from the posterior distribution before estimating the score of each candidate action. This model provides an interpretable baseline while naturally representing predictive uncertainty.

%%%%%%%%%%%%%%%%%%%%%%%%%%%%%%%%%%
\subsubsection{Bootstrap Tree Ensemble (BTE)}
The second learner estimates policy scores using an ensemble of bootstrapped regression trees. The ensemble captures nonlinear interactions between network conditions, operator objectives, and provisioning performance. Prediction variability across individual trees is used as an uncertainty estimate for Thompson-style action selection. Together, the two learners represent complementary assumptions regarding the relationship between operational context and provisioning performance.

\section{Results}
\label{sec:Results}
%The contextual-bandit models are trained using the dataset described in Section~\ref{subsec:dataset}. Training, validation, and testing are performed using chronological partitions to preserve the temporal ordering of network observations. During evaluation, the learners receive only the information available before the target control window. After selecting a provisioning policy, the corresponding expert executes the complete provisioning process for that window, and the resulting network performance is evaluated using the objective-aware score function. The experimental comparison includes both contextual-bandit selectors together with the individual provisioning experts acting as static baselines.

The proposed orchestration framework was evaluated using the service-chain provisioning environment described in Section~\ref{sec:methodology}. For every target control window, the contextual-bandit selector received the pre-decision context and selected one provisioning policy map from the available expert portfolio. The selected policy was then executed for the complete target window, and the resulting network performance was evaluated using the objective-aware score defined in Section~\ref{subsec:reward}. The proposed learners were compared against the individual provisioning experts operating as static policies and against a deterministic rule-based orchestrator. All methods were evaluated under the same traffic scenarios, operator objectives, and network conditions to ensure a fair comparison.

\subsection{Policy Selection Behavior}

\begin{figure*}[htbp]
    \centering
    \includegraphics[width=0.75\textwidth]{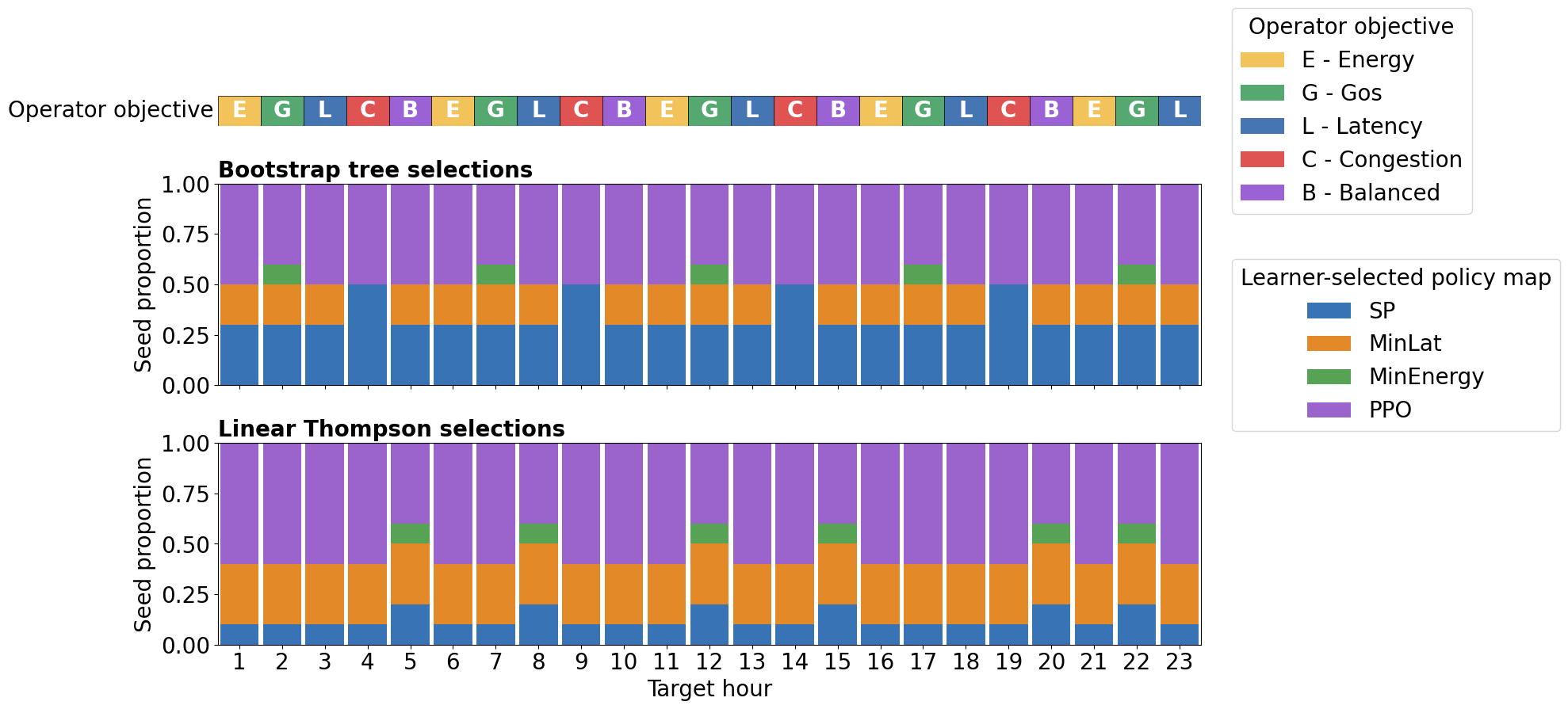}
    \caption{Expert-selection proportions across decision seeds
by target hour. The upper strip identifies the active
operator objective.}
    \label{fig:policy_selection}
\end{figure*}

Figure \ref{fig:policy_selection} shows that the two contextual-bandit learners exhibit markedly different policy-selection behaviors. The \gls{BTE} selector concentrates its decisions on the \gls{SP} and PPO policy while still activating the remaining experts under specific operating conditions. Although PPO and  \gls{SP} remains the dominant selection, the learner consistently switches to the \gls{MinLat} or \gls{MinEnergy} experts depending on the observed context and operator objective. This indicates that the tree-based model identifies operating regions in which different provisioning strategies become preferable.

In contrast, \gls{LT} distributes its decisions more evenly across the available policy maps. The learner frequently alternates between PPO and \gls{MinLat} while assigning a smaller proportion of decisions to \gls{SP} and \gls{MinEnergy}. The relatively balanced action distribution suggests that the linear reward model is less confident in identifying a dominant expert and therefore maintains greater exploration across the available portfolio.

Overall, Figure \ref{fig:policy_selection} demonstrates that the two contextual-bandit models learn substantially different decision policies despite being trained on the same contextual information and objective-aware score. The \gls{BTE} tends to exploit a smaller subset of experts, whereas \gls{LT} maintains a more diversified policy selection strategy.

\subsection{Deterministic Rule-Based Orchestrator}
As a deterministic baseline, a rule-based orchestrator was implemented using a fixed sequence of operational checks and expert-selection rules. The decision process first evaluates operational constraints, such as policy reconfiguration restrictions and explicit operator overrides. It then applies a sequence of safety gates to detect critical service degradation, resource saturation, or forecasted overload conditions. When any safety condition is activated, the orchestrator selects a predefined protective provisioning policy designed to recover or maintain network stability.

If no safety mechanism is triggered, the decision is driven by the operator objective. Latency-oriented objectives select the minimum-latency expert, energy-oriented objectives select the minimum-energy expert, while all remaining situations default to the \gls{MinLat} provisioning policy. Unlike the contextual-bandit approaches, the deterministic orchestrator follows the same decision sequence whenever the same conditions are observed. 
Consequently, it serves as a strong 
%engineering 
baseline that incorporates expert knowledge and operational safety rules but cannot adapt its decision policy from historical observations.

\subsection{Overall Performance}
Table \ref{tab:overall_results} summarizes the aggregate network performance obtained by the evaluated orchestration strategies. For a nonempty window containing $N_t$ requests,
$F_t=N_t-N_t^{\mathrm{acc}}$ counts unsuccessful provisioning
requests and $G_t=N_t^{\mathrm{acc}}/N_t$ is GoS. Among all evaluated methods, the \gls{BTE} selector achieved the highest objective-aware score while simultaneously obtaining the highest \gls{GoS}. These results indicate that learning to select among heterogeneous provisioning experts can provide measurable improvements over relying on a single static provisioning policy. Interestingly, the \gls{BTE} does not optimize every individual metric independently. Instead, it achieves the best overall objective-aware performance by selecting different experts according to the current operating context, illustrating the benefit of policy-level orchestration.

\begin{table}[htbp]
    \centering
    \caption{Performance Comparison (average across evaluation runs).}
    \label{tab:overall_results}
  \resizebox{\columnwidth}{!}{ 
    \begin{tabular}{l c c c c c}
        \hline
        \textbf{Method} & \textbf{Score} & \textbf{GoS} & \textbf{Energy} & \textbf{Latency} & \textbf{Failures} \\
         &  &  & \textbf{(Wh/window)} & \textbf{(ms)} & \textbf{/window} \\
        \hline
        Static SP & 0.7199 & 0.9526 & 17,048 & 4.739 & 94.99 \\
        Bootstrap tree (BTE) & 0.7513 & 0.9710 & 17,313 & 4.929 & 90.17 \\
        Deterministic & 0.7050 & 0.9597 & 16,783 & 5.007 & 120.81 \\
        Static PPO & 0.7108 & 0.9482 & 16,751 & 5.043 & 103.86 \\
        Linear Thompson (LT) & 0.6846 & 0.9360 & 16,706 & 4.848 & 128.32 \\
        Static MinLat & 0.6739 & 0.9501 & 17,234 & 4.607 & 81.93 \\
        Static MinEnergy & 0.6537 & 0.8599 & 14,883 & 4.895 & 280.65 \\
        \hline
    \end{tabular}
    }
\end{table}
% ---
The deterministic rule-based orchestrator achieves competitive energy consumption while maintaining acceptable \gls{GoS}. However, because its decisions are entirely rule-driven, it lacks the ability to adapt its policy-selection strategy from accumulated operational experience. Consequently, its overall objective-aware score remains below that of the learning-based \gls{BTE} selector. On the other hand, \gls{LT} obtains the lowest performance among the orchestration approaches despite employing the same contextual information as the \gls{BTE}. This result suggests that the relationship between network context and expert performance cannot be adequately represented by a globally linear reward model. In contrast, the nonlinear tree ensemble is better able to capture conditional interactions between operator objectives, traffic conditions, and network state.

The comparison with the static provisioning policies further illustrates the motivation for policy-level orchestration. While each individual expert performs well for particular metrics—for example, \gls{MinEnergy} minimizes energy consumption and \gls{MinLat} achieves the lowest average latency—none consistently provides the highest objective-aware score across all operating conditions. The proposed orchestration framework therefore enables the network to exploit the strengths of multiple provisioning experts without modifying their internal optimization algorithms.

\section{Conclusions and Future Work}
\label{sec:Conclusion}

This paper introduced an objective-aware orchestration framework for \gls{O-RAN} service-chain provisioning that operates at the policy level rather than directly optimizing individual routing or resource-allocation decisions. The proposed architecture selects one provisioning policy map for each control window based on the current operator objective and network context, while leaving request-level provisioning to existing deterministic or learning-based experts.

Two contextual bandit selectors with static provisioning policies and a rule-based deterministic orchestrator were investigated. The evaluation framework further introduced an objective-aware score that explicitly prioritizes safe network operation while accommodating different operational objectives. The experimental results demonstrate that learning-based policy orchestration is a promising alternative to static expert selection, allowing different provisioning strategies to be activated according to the operating conditions without modifying or retraining the underlying provisioning algorithms.

Future work will focus on extending the current framework in several directions. First, the present implementation selects a single policy map for the entire network; future work will investigate service-specific and region-specific policy orchestration. Second, the current evaluation is based on canonical one-step counterfactual execution, whereas future work will consider fully closed-loop orchestration in which each selected policy influences subsequent network states. Finally, the proposed contextual-bandit selector will be integrated into a complete agentic orchestration framework in which large language models translate operator intent into structured objectives while the learning component remains responsible for objective-aware policy selection.

\section*{Acknowledgment}
This work was supported by NSERC (under project ALLRP 566589-21) and InnovÉÉ (INNOV-R program) through the partnership with Ericsson. We are grateful to Adel Larabi at GAIA, Ericsson Montréal for clarifying some concepts of the current 5G technology.

% =======================================================
% =======================================================
% =======================================================
\bibliographystyle{IEEEtran}
\bibliography{IEEEabrv, Biblio/Introduction, Biblio/Literature_review, Biblio/Problem, Biblio/BJ}
\end{document}